\documentclass[preprint,review,12pt]{elsarticle}

\usepackage{graphicx}
\usepackage{amssymb}
\usepackage{xcolor}
\usepackage[labelfont=bf]{caption}

\def\etal{\emph{et al.~}}

\journal{Frontiers of Medicine}

\begin{document}

\begin{frontmatter}


\title{Artificial Intelligence in Surgery}



\author{Xiao-Yun Zhou$^1$, Yao Guo$^1$, Mali Shen$^1$, Guang-Zhong Yang$^{2}$}

\address{1. The Hamlyn Centre for Robotic Surgery, Imperial College London, UK \\
2. Institute of Medical Robotics, Shanghai Jiao Tong University, China}

\begin{abstract}
Artificial Intelligence (AI) is gradually changing the practice of surgery with the advanced technological development of imaging, navigation and robotic intervention. In this article, the recent successful and influential applications of AI in surgery are reviewed from pre-operative planning and intra-operative guidance to the integration of surgical robots. We end with summarizing the current state, emerging trends and major challenges in the future development of AI in surgery. 
\end{abstract}


\end{frontmatter}
%
%
%

\section{Introduction}
Advances in surgery have made a significant impact on the management of both acute and chronic diseases, prolonging life and continuously extending the boundary of survival. These advances are underpinned by continuing technological developments in diagnosis, imaging, and surgical instrumentation. Complex surgical navigation and planning are made possible through the use of both pre- and intra-operative imaging techniques such as ultrasound, Computed Tomography (CT), and Magnetic Resonance Imaging (MRI)~\cite{vitiello2012emerging}; surgical trauma is reduced through Minimally Invasive Surgery (MIS), now increasingly combined with robotic assistance~\cite{troccaz2019frontiers}; post-operative care is also improved by sophisticated wearable and implantable sensors for supporting early discharge after surgery, enhancing the recovery of patients and early detection of post-surgical complications~\cite{yang2014body,yang2018implantable}. Many terminal illnesses have been transformed into clinically manageable chronic lifelong conditions and increasing surgery is focused on the systematic level impact on patients, avoiding isolated surgical treatment or anatomical alteration, with careful consideration of metabolic, haemodynamic and neurohormonal consequences that can influence the quality of life.  

For recent advances in medicine, AI has played an important role in clinical decision support since the early years of developing the MYCIN system~\cite{shortliffe2012MYCIN}. AI is now increasingly used for risk stratification, genomics, imaging and diagnosis, precision medicine, and drug discovery. The introduction of AI in surgery is more recent and it has a strong root in imaging and navigation, with early techniques focused on feature detection and computer assisted intervention for both pre-operative planning and intra-operative guidance. Over the years, supervised algorithms such as active shape models, atlas based methods and statistical classifiers have been developed~\cite{vitiello2012emerging}. With recent successes of AlexNet~\cite{krizhevsky2012imagenet}, deep learning methods, especially Deep Convolutional Neural Network (DCNN) where multiple convolutional layers are cascaded, have enabled automatically learned data-driven descriptors, rather than ad hoc hand-crafted features, to be used for image understanding with improved robustness and generalizability.  
With increasing use of robotics in surgery, AI is set to transform the future of surgery, through the development of more sophisticated sensorimotor functions with different levels of autonomy that can give the system the ability to adapt to constantly changing and patient-specific \textit{in vivo} environment, leveraging the parallel advances in medicine in early detection and targeted therapy~\cite{yang2017medical}. It is reasonable to expect that future surgical robots would be able to perceive and understand complicated surroundings, conduct real-time decision making and perform desired tasks with increased precision, safety, and efficiency. But what are the roles of AI in these systems and the future of surgery in general? How to deal with dynamic environments and learn from human operators? How to derive reliable control policy and achieve human-machine symbiosis? 

\begin{figure}[ht]
\centering
\includegraphics[width = \hsize]{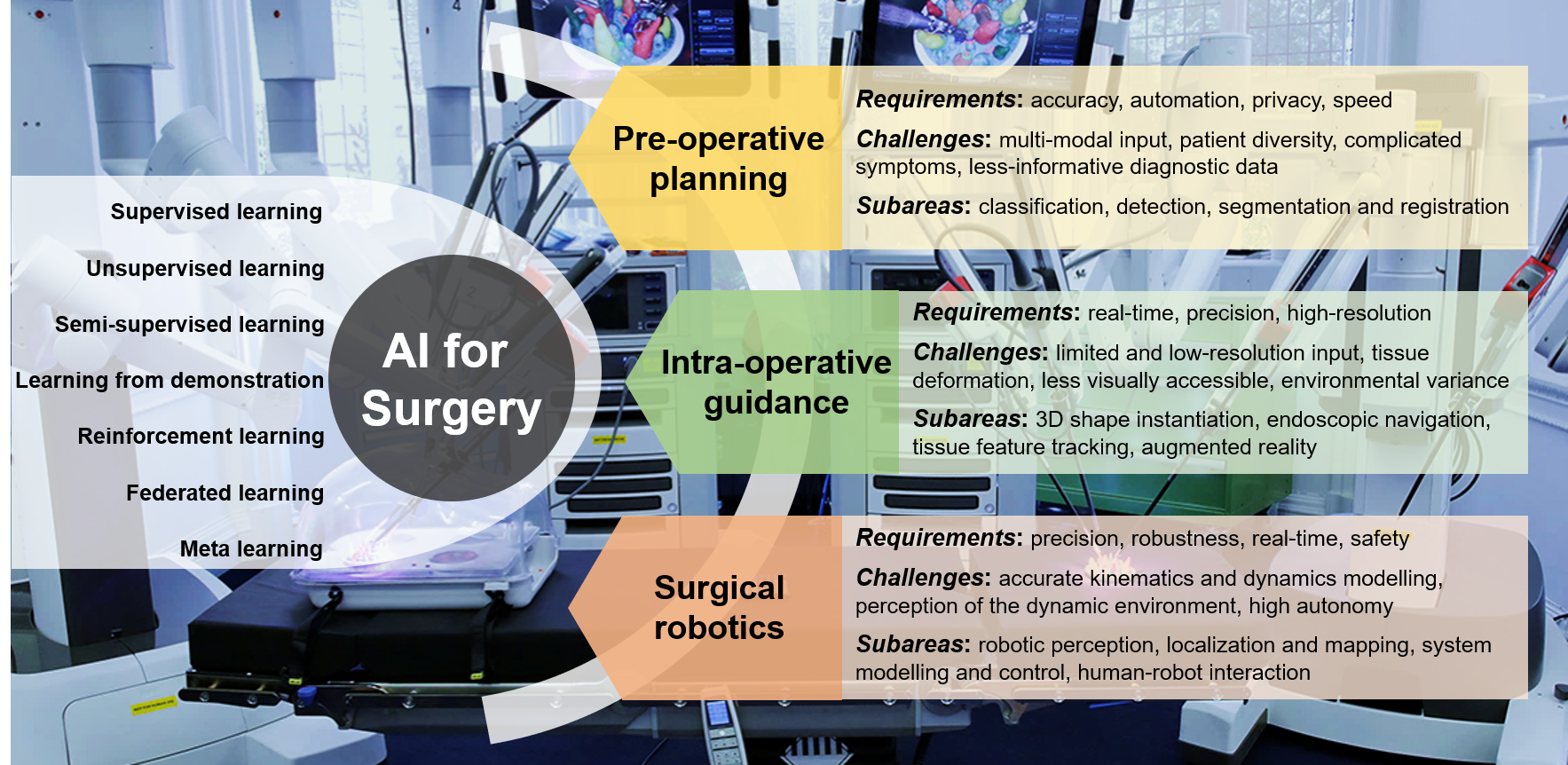}
\caption{An overview of popular AI techniques, as well as the key requirements, challenges, and subareas of AI used in pre-operative planning, intra-operative guidance and surgical robotics.}
\label{fig:AIinro}
\end{figure}

In this article, we review the applications of AI in pre-operative planning, intra-operative guidance, as well as its integrated use in surgical robotics. Popular AI techniques including an overview of their requirements, challenges and subareas in surgery are outlined in Fig.~\ref{fig:AIinro}, showing the main flow of the contents of the paper. We first introduce the application of AI in pre-operative planning and this is followed by AI techniques for intra-operative guidance, a review of AI in surgical robotics, as well as conclusions and future outlook. Technically, we put a strong emphasis on deep learning based approaches in this review. 

\section{AI for Pre-operative Planning}
Pre-operative planning where surgeons plan the surgical procedure from existing medical records and imaging is essential for the success of a surgery. Among existing imaging modalities, X-ray, CT, ultrasound and MRI are the most common ones used in practice. Routine tasks based on medical imaging include anatomical classification, detection, segmentation, and registration.

\subsection{Classification}

Classification outputs the diagnostic value of the input which is a single or a set of medical images or volumes of organs or lesions. In addition to traditional machine learning and image analysis techniques, deep learning based methods for pre-operative planning are on the rise \cite{litjens2017survey}. For the latter,  the network architecture for classification is composed of convolutional layers for extracting information from the input images or volumes and fully connected layers for regressing the diagnostic value. 

For example, a classification pipeline with a Convolutional Neural Network (CNN) architecture of Google's Inception, with Inception and ResNet algorithm and with different training strategies has been proposed to segment the lung, bladder and breast cancer types \cite{khosravi2018deep}. Chilamkurthy \etal demonstrate that deep learning can recognize intracranial haemorrhage, calvarial fracture, midline shift and mass effect through testing a set of deep learning algorithms on head CT scans \cite{chilamkurthy2018deep}. The mortality, renal failure and post-operative bleeding in patients after cardiosurgical care can be predicted by Recurrent Neural Network (RNN) in real time with improved accuracy compared to standard-of-care clinical tools \cite{meyer2018machine}. ResNet-50 and Darknet-19 have been used to classify benign or malignant lesions in ultrasound images, showing similar sensitivity and improved specificity  \cite{li2019diagnosis}. Many studies show promising human-level accuracy with good reproducibility, but explainability of these approaches remains a potential hurdle for regulatory considerations.

\subsection{Detection}
Detection provides the spatial localization of regions of interest, often in the form of bounding boxes or landmarks, additionally to image- or region-level classification. Similarly, deep learning based approaches have shown promises. Compared to traditional algorithms which are task-specific due to hand-crafted feature extractors, DCNNs for detection usually consist of convolutional layers for feature extraction and regression layers for regressing the bounding box properties.

For detecting prostate cancer from 4D Positron-Emission Tomography (PET) images, a deeply stacked convolutional autoencoder was trained to extract the statistical and kinetic biological features \cite{rubinstein2019unsupervised}. For pulmonary nodule detection, 3D CNNs with roto-translation group convolutions (3D G-CNNs) were proposed with good accuracy, sensitivity and convergence speed \cite{winkels2019pulmonary}. CNNs are frequently used in orthopaedics for cartilage lesion detection \cite{liu2018deepZHOU}. For breast lesion detection, Deep Reinforcement Learning (DRL) based on an extension of the deep Q-network was used to learn a search policy from dynamic contrast-enhanced MRI \cite{maicas2017deep}. To detect acute intracranial haemorrhage from CT scans and to improve network interpretability, Lee \etal \cite{lee2019explainable} used an attention map and an iterative process to mimic the workflow of radiologists.

\subsection{Segmentation}
Segmentation can be treated as a pixel- or voxel-level image classification problem. Due to the limitation of computational resources, early works on deep learning for segmentation often adopted a sliding window-based system. Specifically, each image or volume was divided into small windows, CNNs were trained to predict the target label at the central location of the window. Image- or voxel-wise segmentation can be achieved by running the CNN classifier over densely sampled image windows. One of the well-known networks that falls into this category is Deepmedic, which had shown good performances for multi-modal brain tumour segmentation from MRI \cite{kamnitsas2017efficient}. However, the sliding window-based system is inefficient as the network activations of overlapping regions were computed repeatedly. More recently, it was replaced by Fully Convolutional Networks (FCNs) \cite{long2015fully}. The key idea was to replace the fully connected layers in a classification network with convolutional layers and up-sampling layers, which significantly improved the segmentation efficiency. For medical image segmentation, U-Net \cite{ronneberger2015u} \cite{cciccek20163d}, or more generally, encoder-decoder network is a representative FCN that has shown promising performances. The encoder has multiple convolutional and down-sampling layers that extract image features at different scales. The decoder has convolutional and up-sampling layers that recover the spatial resolution of feature maps and finally achieves pixel- or voxel-wise dense segmentation. A review of different normalization methods in training U-Net for medical image segmentation could be found in \cite{zhou2019normalization}.

For navigating the endoscopic pancreatic and biliary procedures, Gibson \etal \cite{gibson2018automatic} used dilated convolutions and fused image features at multiple scales for segmenting abdominal organs from CT scans. For interactive segmentation of placenta and fetal brains from MRI, FCN and user defined bounding boxes and scribbles were combined, where the last few layers of FCN were fine-tuned based on the user input \cite{wang2018interactive}. For aortic MRI, Bai \etal \cite{bai2018recurrent} combined FCN with RNN to incorporate spatial and temporal information. The segmentation and localization of surgical instrument landmarks were modelled as heatmap regression and  FCN was used to track the instruments in near real-time \cite{laina2017concurrent}. For the segmentation and labelling of vertebrae from CT and MRI, Lessmann \etal proposed an iterative instance segmentation approach with FCN, where the network concurrently performed vertebra segmentation, regressed the anatomical landmark and predicted the vertebrae visibility \cite{lessmann2019iterative}. For pulmonary nodule segmentation, Feng \etal addressed the issue of requiring accurate manual annotations when training FCNs by learning discriminative regions from weakly-labelled lung CT with a candidate screening method \cite{feng2017discriminative}.

\subsection{Registration}
Registration is the spatial alignment between two medical images, volumes or modalities, which is particularly important for both pre- and intra-operative planning. Traditional algorithms usually iteratively calculate a parametric transformation, i.e., elastic, fluid or B-spline model to minimize a given metric, i.e., mean square error, normalized cross correlation, or mutual information, between the two medical images, volumes or modalities. Recently, deep regression models have been used to replace the traditional time consuming and optimization based registration algorithm.

Example deep learning based approaches include VoxelMorph based on CNN structures for maximizing the standard image matching objective functions by leveraging auxiliary segmentation to map an input image pair to a deformation field \cite{balakrishnan2019voxelmorph}. An end-to-end deep learning framework was proposed with three stages: affine transform prediction, momentum calculation and non-parametric refinement to combine affine registration and vector momentum-parameterized stationary velocity field for 3D medical image registration \cite{shen2019networks}. Pulmonary CT images were registered by training a 3D CNN with synthetic random transformation \cite{eppenhof2018pulmonary}. A weakly supervised framework was proposed for multi-modal image registration, with training on images with higher-level correspondence, i.e., anatomical labels, rather than voxel-level transformation for predicting the displacement field \cite{hu2018weakly}. Markov decision process with each agent trained with dilated FCN was applied to align a 3D volume to 2D X-ray images \cite{miao2018dilated}. BIRNet was proposed to predict deformation from image appearance for image registration, with training an FCN with both the ground truth and image dissimilarity measures, where the FCN was improved with hierarchical loss, gap filling and multi-source strategies \cite{fan2019birnet}. A Deep Learning Image Registration (DLIR) framework was proposed to train CNN on image similarity between fixed and moving image pairs, hence affine and deformable image registration can be achieved in an unsupervised manner \cite{de2019deep}. RegNet has been proposed by considering multi-scale contexts and is trained on artificially generated Displacement Vector Field (DVF) to achieve a non-rigid registration  \cite{sokooti2017nonrigid}. 3D image registration can also be formulated as a strategy learning process with 3D raw image as the input, the next optimal action, i.e., up and down, as the output, CNN as the agent \cite{liao2017artificial}.
\section{AI for Intra-operative Guidance}
\begin{figure}[ht]
\centering
\includegraphics[width = \hsize]{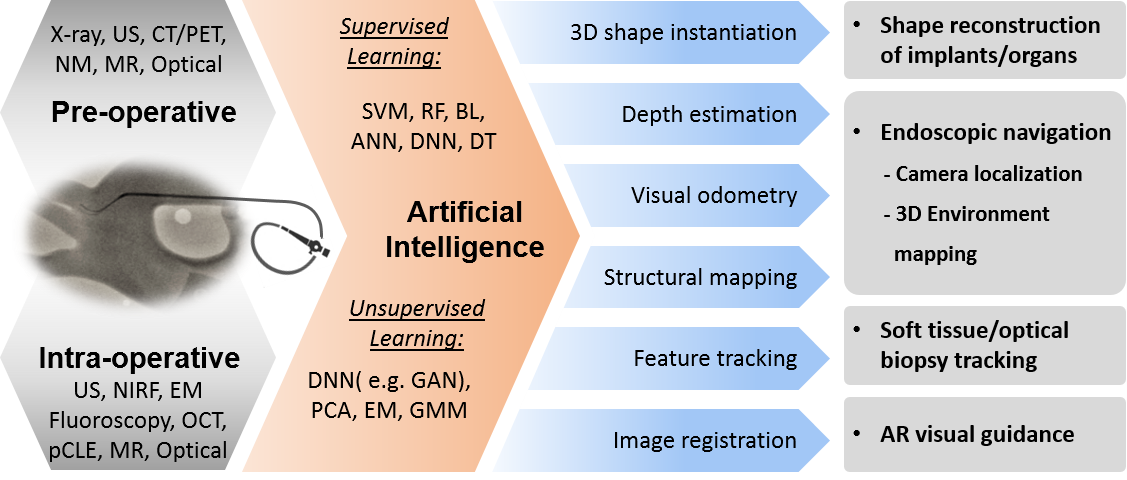}
\caption{AI techniques for computer-aided intra-operative guidance in MIS. Multi-modal data acquired pre-operatively and intra-operatively are used in either supervised or unsupervised learning based techniques for various surgical applications (US - ultrasound; NIRF - near infrared fluorescence ; OCT - optical coherence tomography; pCLE - probe-based confocal laser endomicroscopy; EM sensor - electromagnetic sensor; RF - random forests; BL - bayesian learning; DT - decision tree; EM - expectation maximization; GMM - gaussian mixture models.)
}
\label{fig:AIforNavigation}
\end{figure}

Computer-aided intra-operative guidance has always been a corner-stone of MIS. Learning strategies have been extensively integrated into the development of intra-operative guidance to provide enhanced visualization and localization in surgery. For the purpose of intra-operative guidance, recent work can be divided into four main aspects: intra-operative shape instantiation, endoscopic navigation, tissue tracking and Augmented Reality (AR), as summarized in Fig. \ref{fig:AIforNavigation}.

\subsection{3D Shape Instantiation}
For intra-operative 3D reconstruction, 3D volumes can be scanned with  MRI, CT or ultrasound. In practice, this process (3D/4D) can be time-consuming or with a low resolution. Real-time 3D shape instantiation which instantiates the intra-operative 3D shape from a single or limited 2D images is an emerging area of research in intra-operative guidance. 

For example, a 3D prostate shape was instantiated from multiple non-parallel 2D ultrasound images with a radial basis function \cite{cool20063d}. The 3D shape of Abdominal Aortic Aneurysm (AAA) was instantiated from two 2D fluoroscopic images \cite{toth2015adaption}. The 3D shapes of fully-compressed, fully-deployed and also partially-deployed stent grafts were instantiated from a single projection of 2D fluoroscopy with mathematical modelling, combined with the Robust Perspective-n-Point (RPnP) method, graft gap interpolation and graph neural networks \cite{zhou2017stent, zhou2018real,zheng2019RT}. Furthermore, equally weighted focal U-Net \cite{zhou2018real} was proposed to automatically segment the makers on stent grafts to improve the efficiency of the intra-operative stent graft shape instantiation framework \cite{zhou2018towards}. Moreover, the 3D AAA skeleton was instantiated from a single projection of 2D fluoroscopy with skeleton deformation and graph matching \cite{zheng20183d}. The 3D liver shape was instantiated from a single 2D projection with Principal Component Analysis (PCA), Statistical Shape Model (SSM) and Partial Least Square Regression (PLSR) \cite{lee2010dynamic}. This work was further generalized to a registration-free shape instantiation framework for any dynamic organ with sparse PCA, SSM and kernel PLSR \cite{zhou2018areal}. Recently, an advanced deep and one-stage learning strategy that estimates 3D point cloud from a single 2D image was proposed for 3D shape instantiation \cite{zhou2019one}. 

\subsection{Endoscopic Navigation}
In surgery, there is an increasing trend towards intra-luminal procedures and endoscopic surgery driven by early detection and intervention. Navigation techniques have been investigated to guide the manoeuvre of endoscopes towards target locations. To this end, learning-based depth estimation, visual odometry and Simultaneous Localization and Mapping (SLAM) have been tailored for camera localization and environment mapping with the use of endoscopic images. 

\subsubsection{Depth estimation}
Depth estimation from endoscopic images plays an essential role in 6 DoF camera motion estimation and 3D structural environment mapping, which has been tackled either by supervised \cite{mahmood2018deep,mahmood2018unsupervised} or by self-supervised~\cite{turan2018unsupervised,shen2019context} deep learning methods.
Compared to natural images of indoor or outdoor scenes, depth recovery from endoscopic images suffer from two main challenges. First, it is practically difficult to obtain a large amount of high-quality training data containing paired video images and depth maps due to both hardware constraints and labour-intensive labelling. Therefore, conventional supervised depth recovery methods such as \cite{liu2015deep} are not applicable in endoscopic image scenarios. Second, surgical scenes are often textureless, making it difficult to apply depth recovery methods that rely on feature matching and reconstruction \cite{Zhou_2017_CVPR,Zhan_2018_CVPR}. 

To address the challenge of limited training data, Ye \etal \cite{ye2017self} proposed a self-supervised depth estimation approach for stereo images using siamese networks. For monocular depth recovery, Mahmood \etal\cite{mahmood2018deep,mahmood2018unsupervised} learnt the mapping from rendered RGB images to the corresponding depth maps with synthetic data and adopted domain transfer learning to convert real RGB images to rendered images. Additionally, a self-supervised unpaired image to image translation \cite{shen2019context} using a modified Cycle Generative Adversarial Network (CycleGAN) \cite{zhu2017unpaired} was proposed to recover the depth from bronchoscopic images. Moreover, a self-supervised CNN based on the principle of Shape from Motion (SFM) was applied to recover the depth and achieve visual odometry for endoscopic capsule robot \cite{turan2018unsupervised}. 
\subsubsection{Visual odometry}
Visual odometry uses consecutive video frames to estimate the pose of a moving camera. CNN-based approaches~\cite{turan2018deep} have been adopted for camera pose estimation based on temporal information. Turan \etal\cite{turan2018deep} estimated the camera pose for endoscopic capsule robot with using CNN for feature extraction and Long Short-Term Memory (LSTM) for dynamics estimation. Sganga \etal\cite{sganga2018offsetnet} combined ResNet and FCN for calculating the pose change between consecutive video frames. However, the feasibility of localization approaches based on visual odometry has only been validated on lung phantom data~\cite{sganga2018offsetnet} and Gastrointestinal (GI) tract data~\cite{turan2018deep}.   
\subsubsection{3D reconstruction and localization}
Due to the dynamic nature of tissues, real-time 3D reconstruction of the environment and localization are vital prerequisites for navigation.  
SLAM is a widely studied research topic in robotics, in which the robot can simultaneously builds the 3D map of surrounding environments and localizes the camera pose in the built map. 
Traditional SLAM algorithms are based on the assumption of a rigid environment, which is in contrast to that found in a typical surgical scene where the deformation of soft tissues and organs is involved, limiting its direct adoption to surgical tasks. 
To tackle this challenge, Mountney \etal\cite{mountney2006simultaneous} first applied the Extended Kalman Filter SLAM (EKF-SLAM) framework~\cite{davison2007monoslam} in MIS with a stereo-endoscope, where the SLAM estimation was compensated with periodic motion of soft tissues caused by respiration~\cite{mountney2010motion}. Grasa \etal\cite{grasa2013visual} evaluated the effectiveness of the monocular EKF-SLAM in hernia repair surgery for measuring hernia defect. In~\cite{turan2017non}, the depth images were first estimated from the RGB data through Shape from Shading (SfS). Then they adopted the RGB-D SLAM framework by using paired RGB and depth images. Song \etal\cite{song2018mis} implemented a dense deformable SLAM on a GPU and a ORB-SLAM on a CPU to boost the localization and mapping performance of a stereo-endoscope. 

Endovascular interventions have been increasingly used to treat cardiovascular diseases. However, visual cameras are not applicable inside vessels, for example, catheter mapping is commonly used in Radiofrequency Catheter Ablation (RFCA) for navigation \cite{zhou2016path}. To this end, recent advances in Intravascular Ultrasound (IVUS) have opened up another avenue for endovascular intra-operative guidance. Shi and Yang first proposed the Simultaneous Catheter and Environment (SCEM) framework for 3D vasculature reconstruction by fusing the Electromagnetic (EM) sensing data and IVUS images  \cite{shi2014simultaneous}. To deal with the errors and uncertainty measured from both EM sensors and IVUS images, the improved SCEM+ solved the 3D reconstruction by solving a nonlinear optimization problem \cite{zhao2016scem}. To further alleviate the burden of pre-registration between pre-operative CT data and EM sensing data, a registration-free SCEM framework \cite{zhao2016registration} was proposed for more efficient data fusion.  

\subsection{Tissue Feature Tracking}
Learning strategies have also been applied to soft tissue tracking in MIS. Mountney \etal\cite{mountney2008soft} introduced an online learning framework that updates the feature tracker over time by selecting correct features using decision tree classification. Ye \etal\cite{ye2016online} proposed a detection approach that incorporates structured Support Vector Machine (SVM) and online random forest for re-targeting a pre-selected optical biopsy region on soft tissue surface of GI tract. Wang \etal\cite{wang20173} adopted a statistical appearance model to differentiate the organ from the background in their region-based 3D tracking algorithm. All their validation results demonstrate that incorporating learning strategies can improve the robustness of tissue tracking with respect to the deformation and illumination variation. 

\subsection{Augmented Reality}
AR improves surgeons' intra-operative vision through a prevision of a semi-transparent overlay of pre-operative imaging on the area of interest. \cite{bernhardt2017status}. Wang \etal\cite{wang2014augmented} used a projector to project the AR overlay for oral and maxillofacial surgery. The 3D contour matching was used to calculate the transformation between the virtual image and real teeth. Instead of using projectors, Pratt \etal exploited Hololens, a head-mounted AR device, to demonstrate the 3D vascular model on the lower limb of patient~\cite{pratt2018through}. While one of the most challenging tasks is to project the overlay on markerless deformable organs, Zhang \etal\cite{zhang2019markerless} introduced an automatic registration framework for AR navigation, of which the Iterative Closet Point (ICP) and RANSAC were applied for 3D deformable tissue reconstruction. 
\section{AI for Surgical Robotics}
\begin{figure}[ht]
\centering
\includegraphics[width = \hsize]{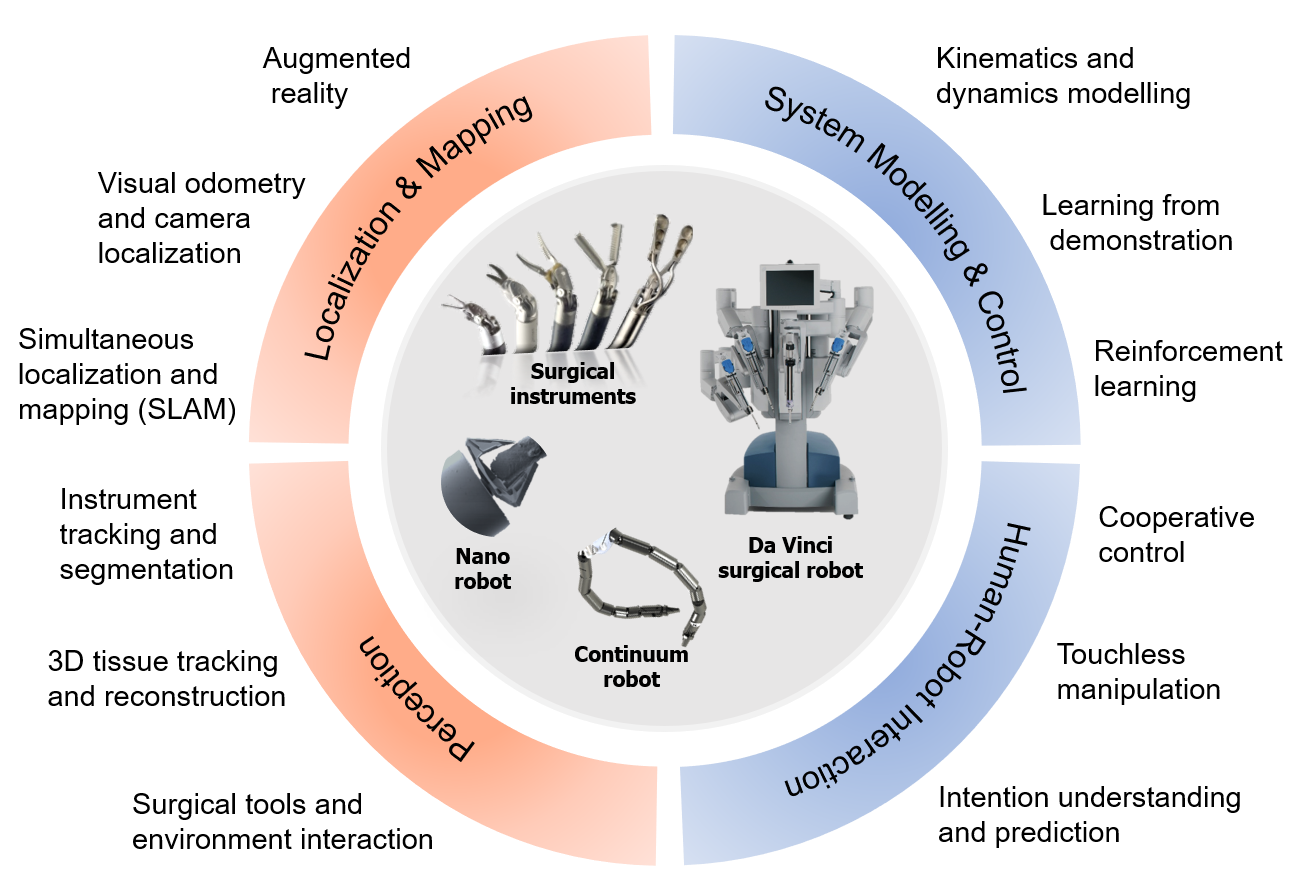}
\caption{AI techniques for surgical robotics including perception, localization \& mapping, system modelling \& control, and human-robot interaction.}
\label{fig:AIforRobot}
\end{figure}
With the development of AI techniques, surgical robots can achieve superhuman performance during MIS \cite{topol2019high,mirnezami2018surgery}. The objective of AI is to boost the capability of surgical robotic systems in perceiving the complex \textit{in vivo} environment, conducting decision making, and performing the desired task with increased precision, safety, and efficiency. 
As illustrated in Fig. \ref{fig:AIforRobot}, common AI techniques used for Robotic and Autonomous Systems (RAS) can be summarized in the following four aspects: 1) perception, 2) localization and mapping, 3) system modelling and control, and 4) human-robot interaction. As overlap exists between intra-operative guidance and robot localization \& mapping, this section mainly covers the methods for increasing the level of autonomy in surgical robotics.

\subsection{Perception}
\subsubsection{Instrument segmentation and tracking}
The instrument segmentation task can be divided into three groups: segmentation for distinguishing the instrument and background, multi-class segmentation of instrument parts, i.e., shaft, wrist, and gripper, and multi-class segmentation for different instruments. 
The advancement of deep learning in segmentation has significantly improved the instrument segmentation accuracy from the exploitation of SVM for pixel-level binary classification \cite{bouget2015detecting} to more recent popular DCNN architectures, e.g., U-Net, TernausNet-VGG11, TernausNet-VGG16, and LinkNet based on ResNet architecture, for both binary segmentation and multi-class segmentation \cite{shvets2018automatic}. To further improve the performance, Islam \etal developed a cascaded CNN with a multi-resolution feature fusion framework \cite{islam2019real}.

Algorithms for solving tracking problems can be summarized into two categories: tracking by detection and tracking via local optimization \cite{sznitman2012unified}. 
Previous works in this field mainly relied on hand-crafted features, such as Haar wavelets \cite{sznitman2012unified}, color or texture features \cite{zhang2017real}, and gradient-based features \cite{ye2016real}. 
These methods have different advantages and disadvantages. 
In the context of deep learning based surgical instrument tracking, the proposed methods were built on the tracking by detection \cite{zhao2017tracking,nwoye2019weakly}. Various CNN architectures, e.g., AlexNet \cite{zhao2017tracking} and ResNet \cite{nwoye2019weakly,laina2017concurrent}, were used for detecting the surgical tools from RGB images while \cite{sarikaya2017detection} additionally fed the optical flow estimated from color images into the network. In order to leverage the spatiotemporal information, the LSTM was integrated to smooth the detection results \cite{nwoye2019weakly}. In addition to the position tracking, the pose of articulated end-effector was simultaneously estimated by the methods in \cite{ye2016real,kurmann2017simultaneous}. 

\subsubsection{Surgical tools and environment interaction}
A representative example of tool-tissue interaction during surgery is suturing. In this task, the robot needs to recover the 2D or 3D shape of thread from 2D images in real-time. 
Other challenges to be addressed for this task include the deformation of thread and variations of the environment. Padoy \etal\cite{padoy20113d} introduced a Markov Random Field (MRF) based optimization method to track the 3D thread modelled by a Non-Uniform Rational B-Spline (NURBS). Recently, a supervised two-branch CNN called Deep Multi-Stage Detection (DMSD), was proposed for surgical thread detection \cite{hu2018multi}. In addition, they improved the DMSD framework with a CycleGAN \cite{zhu2017unpaired} structure for the foreground and background adaptation \cite{gu2018cross}. Based on adversarial learning, more synthetic data for thread detection was generated while preserving the semantic information, which enabled the learned knowledge to be transferred to the target domain. 

The estimation of the interaction force between surgical instruments and tissues can provide meaningful feedbacks to ensure a safe manipulation. Due to the limited size of surgical tools for MIS, the high precision and miniaturized force sensors are still immature. Recent works have incorporated AI techniques in the field of Vision-based Force Sensing (VBFS), which can accurately estimate the force values from visual inputs. The LSTM-RNN architecture can automatically learn the accurate mapping between visual-geometric information and applied force in a supervised manner \cite{aviles2016towards}. In addition to the supervised learning, a semi-supervised DCNN was proposed in \cite{marban2018estimation}, where the convolution auto-encoder learns the representation from RGB images followed by minimizing the error between the estimated force and ground truth using the LSTM. 

\subsection{System Modelling and Control}
\subsubsection{Learning from human demonstrations}
Learning from demonstration (LfD), also known as programming by demonstration, imitation learning, and apprenticeship learning, is a popular paradigm for enabling robots to autonomously perform new tasks with the learned policies.
This paradigm is beneficial for complicated automation tasks such as surgical procedures, for which surgical robots can autonomously execute specific motions or tasks simply through learning from surgeons' demonstrations without tedious programming procedures. The robots could reduce surgeons' tedium as well as providing superhuman performance both fast speed and smoothness. The common framework of LfD is to first segment a complicated surgical task into several motion primitives or subtasks, followed by recognition, modelling and execution of these motion primitives sequentially. 

\paragraph{A. Surgical task segmentation and recognition}\mbox{}\\
\indent JHU-ISI Gesture and Skill Assessment Working Set (JIGSAWS) dataset \cite{ahmidi2017dataset} is the first public benchmark dataset for surgical activity segmentation and recognition. This dataset contains the synchronized video and kinematic data (3D motion trajectory and 3D rotation of the end-effector) of three subtasks captured from the Da Vinci robot: suturing, needle passing, and knot tying. For surgical task segmentation, unsupervised clustering algorithms are most popular. In \cite{fard2016soft}, a soft boundary modified Gath-Geva clustering was proposed for segmenting kinematic data. A Transition State Clustering (TSC) method \cite{krishnan2017transition} was presented to exploit both the video and kinematic data to detect and cluster transitions between linear dynamic regimes based on kinematic, sensory and temporal similarity. The authors extended their TSC method to improve the segmentation results by applying DCNNs for extracting features from video data \cite{murali2016tsc}. For surgical subtask recognition, most previous works \cite{ahmidi2017dataset,zappella2013surgical,tao2013surgical} were developed towards variations on Hidden Markov Model (HMM), Conditional Random Field (CRF), and Linear Dynamic Systems (LDS). Particularly, the joint segmentation and recognition frameworks were proposed in \cite{despinoy2015unsupervised,dipietro2019segmenting}. In specific, \cite{dipietro2019segmenting} modelled complex and non-linear dynamics of kinematic data with RNN to recognize both surgical gestures and activities. They compared the simple RNN, forward LSTM, Bidirectional LSTM, Gated Recurrent Unit (GRU), and Mixed history RNN with traditional methods in terms of surgical activity recognition. Liu \etal \cite{liu2018deep} introduced a novel method by modelling the recognition task as a sequential decision-making process and trained an agent by Reinforcement Learning (RL) with hierarchical features from a DCNN model. 

\paragraph{B. Surgical task modelling, generation, and execution}\mbox{}\\
\indent After acquiring the segmented motion trajectories representing surgical subtasks, e.g., suturing, needle passing, and knot tying, the Dynamic Time Warping (DTW) algorithm can be applied to temporally align different demonstrations before modelling. In order to autonomously generate the motion in a new task, Gaussian Mixture Model (GMM)~\cite{padoy2011human,calinon2014human}, Gaussian Process Regression (GPR)~\cite{osa2014online}, dynamics model~\cite{van2010superhuman}, finite state machine \cite{murali2015learning}, and RNN~\cite{mayer2008system} were extensively studied for modelling the demonstrated trajectories in previous works.   
The experts' demonstrations are encoded by the GMM algorithm, and the parameters of mixture model can be iteratively estimated by the expectation maximization algorithm. With the given GMM, the Gaussian Mixture Regression (GMR) was then used to generate the target trajectory of the desired surgical task~\cite{padoy2011human,calinon2014human}. GPR is a non-linear Bayesian function learning technique that models a sequence of observations generated by a Gaussian process. Osa \etal\cite{osa2014online} chose GPR for online path planning in a dynamic environment. 
Given the predicted motion trajectory, different control strategies, e.g., Linear-Quadratic Regulator (LQR) controller \cite{van2010superhuman}, sliding mode control \cite{osa2014online}, neural network \cite{de2016neural}, etc., can be applied to improve the robustness in surgical task execution. 

\subsubsection{Reinforcement learning}
In many surgical tasks, RL is another popular machine learning paradigm to solve the problem that is difficult to analytically model and explicitly observe~\cite{kober2013reinforcement}, e.g., control of the continuum robot, soft tissue manipulation, cutting gauze tensioning, tube insertion, etc.. In the learning process, the controller of autonomous surgical robot, known as an agent, tries to find the optimized policies that yield highly accumulated reward through iterative interaction with the surrounding environment. The environment of RL is modelled as a Markov Decision Process (MDP). To efficiently reduce the learning time, the RL algorithm can be initialized with the learned policies from human expert demonstrations \cite{abbeel2004apprenticeship,calinon2014human,tan2019robot}.  Instead of learning from scratch, the robot can improve the initial policy based on the demonstrations to reproduce the desired surgical tasks. In \cite{tan2019robot}, a Generative Adversarial Imitation Learning (GAIL) \cite{ho2016generative} agent was trained to imitate latent patterns existed in human demonstrations, which can deal with the mismatch distribution caused by multi-modal behaviours. Recently, DRL with advanced policy search methods endows robots to autonomously execute a wide range of tasks \cite{levine2016end}. However, it is unrealistic to repeat the experiments on the surgical robotic platform for over a million times. To this end, the agent can be first trained in a simulation environment and transferred to a real robotic system.  \cite{thananjeyan2017multilateral} first learned tensioning policies from a finite-element simulator via DRL, and then transferred to a real physical system. However, the discrepancy between the simulation data and the real-world environment remains less developed.  

\subsection{Human-Robot Interaction}
Human-Robot Interaction (HRI) is a field that integrates knowledge and techniques from multiple disciplines to build an effective communication between human and robots. With the help of AI, surgical task-oriented HRI allows surgeons to cooperatively control the surgical robotic systems with touchless manipulation. Interaction mediums between surgeons and intelligent robots are usually through surgeons' gaze, head movement, speech/voice, and hand gesture. By understanding the intention of human, robots can then perform the most appropriate actions that satisfy surgeons' needs.  

The 2D/3D eye-gaze point of surgeons tracked via head-mounted or remote eye trackers can assist surgical instrumental control and navigation \cite{yang2002visual}. For surgical robots, the eye-gaze contingent paradigm is able to facilitate the transmission of images and enhance the procedure performance, enabling much more accurate navigation of the instruments \cite{yang2002visual}. Yang \etal\cite{yang2008perceptual} first introduced the concept of gaze-contingent perceptual docking for robot-assisted MIS in 2008, in which the robot can learn the operators' specific motor and perceptual behaviour through their saccadic eye movements and ocular vergence. Inspired by this idea, Visentini \etal\cite{visentini2009brush} used the gaze-contingent to reconstruct the surgeon's area of interest with a Bayesian chains method in real-time. Fujii \etal\cite{fujii2018gaze} performed gaze gesture recognition with the HMM so as to pan, zoom, and tilt the laparoscope during the surgery. In addition to the use of human gaze, the recognition of surgeons' head movement can also be used to control laparoscope or endoscope remotely \cite{nishikawa2003face,hong2019head}. 

Robots have the potential to interpret humans' intentions or commands through voice commands, but for assisting robotic surgery, it still remains challenging due to the noisy environment 
in the operation room. With the development of deep learning in speech recognition, the precision and the accuracy of speech recognition have been significantly improved \cite{graves2013speech}. This improvement leads to a more reliable control of the surgical robot \cite{zinchenko2016study}. 

Moreover, hand gesture is another popular medium in different HRI scenarios. In the previous works, learning-based real-time hand gestures detection and recognition methods have been studied by taking advantages of different sensors. Jacob \etal\cite{jacob2012gestonurse,jacob2013collaboration} designed a robotic scrub nurse, Gestonurse, to understand nonverbal hand gestures. They used the Kinect sensor to localize and recognize different gestures generated by surgeons, which can help to deliver surgical instruments to surgeons. Wen \etal introduced an HMM-based hand gesture recognition method for AR control \cite{wen2014hand}, and more recently, with the help of deep learning, vision-based hand gesture recognition with high precision \cite{oyedotun2017deep} can be achieved, therefore, significantly improve the safety for HRI in surgery. 
\section{Conclusion and Future Outlook}
\begin{figure}[ht]
\centering
\includegraphics[width = \hsize]{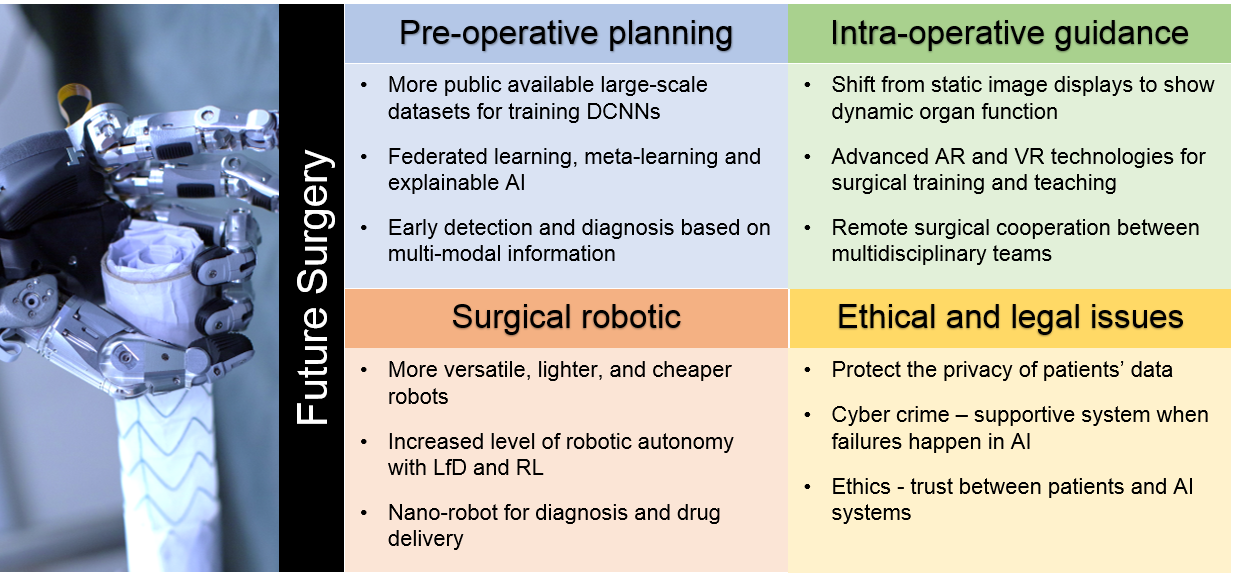}
\caption{An outlook of the future of surgery in pre-operative planning, intra-operative guidance, surgical robotics, and also potentially caused ethical and legal issues.}
\label{fig:outlook}
\end{figure}
The advancement in AI has been transforming modern surgery towards more precise and autonomous intervention for treating both acute and chronic symptoms. By leveraging such techniques, marked progresses have been made in pre-operative planning, intra-operative guidance and surgical robotics. In the following, we summarize the major challenges for these three aspects as shown in Fig. \ref{fig:outlook}, and then discuss achievable visions of the future directions. Finally, other key issues, such as ethics, regulation, and privacy, are further discussed.

\subsection{Pre-operative Planning}
Deep learning has been widely adopted in pre-operative planning for tasks ranging from anatomical classification, detection, segmentation to image registration. The results seem to suggest that the deep learning based methods can outperform those rely on conventional approaches. However, data-driven approaches often suffer from inherited limitations, making the deep learning based approaches less generalizable, explainable and more data-demanding.

To overcome these issues, close collaborations between multidisciplinary teams, particularly the surgeons and AI researchers should be encouraged to generate large scale annotated data, providing more training data for AI algorithms. An alternative solution is to develop AI techniques such as meta-learning, or learning to learn, that enable generalizable systems to perform diagnosis with limited dataset yet improved explainability.

Although many state-of-the-art machine learning and deep learning algorithms have made breakthroughs in the field of general computer vision, the differences between medical and natural images can be significant, which may impede their clinical applicability. In addition, the underlying models and the derived results may not be easily interpretable by humans, therefore it raises issues such as potential risks and uncertainty in surgery. Potential solutions to these problems would be to explore different transfer learning techniques to mitigate the differences between image modalities and to develop more explainable AI algorithms to enhance its decision-making performance.

Furthermore, utilizing personalized multimodal patient information, including omics-data and life style information, in the development of AI can be useful in early detection and diagnosis, leading to personalized treatment. These also allow early treatment options featured with minimal trauma, smaller surgical risks and shorter recovery time.

\subsection{Intra-operative Guidance}
AI techniques have already contributed to more accurate and robust intra-operative guidance for MIS. 3D shape instantiation, camera pose estimation and dynamic environment tracking and reconstruction have been tackled to assist various surgical interventions.

For developing computer-assisted guidance from visual observations, key focuses should be on improving the localization and mapping performance with textureless surfaces, variation in illumination, and limited field of view.

Another key challenge is that the deformation of organs/tissues forcing the pre-operative planning to work with a dynamic and uncertain environment during surgery. Although AI technologies have succeeded in detection, segmentation, tracking, and classification, the studies on extending to more sophisticated 3D applications are required. Additionally, during a surgery, one important requirement is to assist surgeons in real-time, and therefore the efficiency of an AI algorithm becomes a crucial issue. Such demands have been encountered in developing AR or VR where frequent interactions are required either between surgeons and autonomous guidance systems or during remote surgery involving multidisciplinary teams located in different geographical locations.

In addition to the visual information, future AI technologies need to fuse multimodal data from various sensors to achieve more precise perception of the complicated environment. Furthermore, the increasing use of micro- and nano-robotics in surgery will come with new guidance issues.

\subsection{Surgical Robotics}
With the integration of AI, surgical robotics would be able to perceive and understand complicated surroundings, conduct real-time decision making and perform surgical tasks with increased precision, safety, automation, and efficiency. For instance, current robots can already automatically perform some simple surgical tasks, such as suturing and knot tying \cite{hu2018robotic,hu2019designing}. Nevertheless, the increased level of robotic autonomy for more complicated tasks could be achieved by advanced LfD and RL algorithms, especially with the consideration of the interaction with dynamic environments. Due to the diversity of surgical robotic platforms, generalized learning for accurate modelling and control is also required.

Most of the current surgical robots are associated with high cost, large size and being only to perform master-slave operations. We believe that a more versatile, lighter and probably cheaper robotic system needs to be developed, so it can access more constrained regions during MIS \cite{troccaz2019frontiers}. Certainly, it also needs to be easily integrated in well-developed surgical workflows, so that the robot can collaborate with the human operators seamlessly. To date, the current technologies in RAS are still far from achieving full autonomy, human supervision would remain to ensure safety and high-level decision making.

In the coming future, intelligent micro- and nano-robots for non-invasive surgeries and drug delivery could be realized. Furthermore, with the data captured during pre-operative examinations, robots could also assist manufacturing personalized 3D bio-printed tissues and organs for transplant
surgery.

\subsection{Ethical and Legal Considerations of AI in Surgery}
Beyond precision, robustness, safety and automation, it is necessary to carefully consider the legal and ethical considerations of AI in Surgery. These include: 1) privacy - patients’ medical records, gene data, illness prediction data, and operation process data need to be protected with high security; 2) cyber crime - impact on patients needs to be minimized when failures happen in AI-based surgical systems which should be verified and certificated while considering all possible risks; 3) ethics – to make sure new technologies are used responsibly, e.g., gene editing and bio-printed organ transplant on long-term human reproduction, and to build the trust between human and AI techniques gradually. 

In conclusion, we still have a long way to go to replicate and match the levels of intelligence that we see in surgeons and “AIs that can learn complex tasks on their own and with a minimum of initial training data will prove critical for next-generation systems”~\cite{yang2018grand}. Here we quote some of the questions raised by Yang \etal in their article on Medical Robotics - Regulatory, Ethical, and Legal Considerations for Increasing Levels of Autonomy \cite{yang2017medical}: \textit{``As the capabilities of medical robotics following a progressive path represented by various levels of autonomy evolve, most of the role of the medical specialists will shift toward diagnosis and decision-making. Could this shift also mean that medical specialists will be less skilled in terms of dexterity and basic surgical skills as the technologies are introduced? What would be the implication on future training and accreditation? … If robot performance proves to be superior to that of humans, should we put our trust in fully autonomous medical robots?"}    Clearly there are many more issues need to be addressed before AI can be more seamless integrated in the future of surgery.
%
%
%
\bibliographystyle{IEEEtran}
\bibliography{Reference}
\end{document}